# How Video Passthrough Headsets Influence Perception of Self and Others

Monique Santoso, BA[1, *] and Jeremy N. Bailenson, PhD[1]

[1]Department of Communication, Stanford University, Stanford, California, USA

*mtsantoso@stanford.edu

**Abstract**
With the increasing adoption of mixed reality headsets with video passthrough functionality, concerns over perceptual and social effects have surfaced. Building on prior qualitative findings,[1] this study quantitatively investigates the impact of video passthrough on users. Forty participants completed a body transfer task twice, once while wearing a headset in video passthrough and once without a headset. Results indicate that using video passthrough induces simulator sickness, creates *social absence,* (another person in the physical room feels less present), alters self-reported body schema, and distorts distance perception. On the other hand, compared to past research which showed perceptual aftereffects from video passthrough, the current study found none. We discuss the broader implications for the widespread adoption of mixed reality headsets and their impact on theories surrounding presence and body transfer.

**Introduction**

Prior to the release of the Meta Quest 3 and Apple Vision Pro mixed reality (MR) headsets, the notion of using a headset over one's field of vision for hours daily seemed unlikely. However, since their launch in early 2023, the companies producing these headsets have commercialized the ability to blend virtual content into the physical world through *video passthrough*, arguing the technology will be able to immerse users while still providing situational awareness.[2,3] Full video color passthrough relies on stereoscopic, high resolution, low latency, and real-time video of the world displayed through headset screens.[4] This is in contrast to see-through displays, wherein users would view content projected onto a transparent lens which allows users to see light from the physical world (for an early comparison of optical and video see-through, review Rolland et al., 1995).[5]

Recent work by Bailenson et al. (2024) on the latest headsets including the Meta Quest 3 and the Apple Vision Pro anecdotally compared user experiences in passthrough and the physical world.[1] The authors of this work found simulator sickness symptoms that ranged from (i) nausea to eye strain within less than an hour of use, (ii) lapses in distance estimations wherein participants would underestimate distances, (iii) social absence, where individuals in the physical world felt less real and distant, as well as (iv) changes in one's body perception and embodiment in passthrough. However, that study did not provide quantitative data. Despite the proliferation of this new technology, the comparison of how passthrough differs from physical face-to-face interactions with one's self and others remains largely underexplored yet increasingly important. As such, the goal of this work is to quantify these prior qualitative findings on simulator sickness, body distortions, social absence and distance underestimation and situate them in psychology and communication theories, as well as the technological affordances of these headsets.

Sensory conflict theory is commonly cited for simulator sickness, wherein conflicting information received from visual, vestibular, and somatosensory systems disrupts the well-trained interaction between senses, therefore, triggering sickness.[6,7] But most simulator sickness studies look at rendering 3D content and scenes, as opposed to real time stereoscopic video of the real world. In a recent study, Guo et al., 2024, found only very minor simulator sickness in passthrough video exergames.[8] Therefore, our first research question aims to address whether short doses of video passthrough can induce simulator sickness (**RQ1**).

Moreover, researchers also noted the changes in individual body perceptions, including the change to their body shape and size in video passthrough compared to without the headset.[1] These changes were attributed to the distortions in the passthrough medium, as well as the technological affordance of camera placement that is higher than the natural eye position. Prior work on body ownership and transfer in augmented reality (AR) have implemented the classic rubber hand experiment, which illustrates that individuals in AR do not report feeling as much ownership of their rubber hand as in the real-world setting.[9] Other work using see-through HMD has illustrated findings that participants do experience body ownership with their mirrored virtual body in a virtual mirror.[10] Despite these, the quantification of embodiment and body ownership, along with how participants experience changes in their perceived body schema in video passthrough and without headset need replication and extension. Thus, our work will quantify the degree to which embodiment and body ownership is experienced in video passthrough when users view their physical bodies before a physical mirror (**RQ2**).

The current study also examines how the perceptual system is impacted while wearing video passthrough and the *aftereffects* which occur once the headset is removed.[11] Previous research has examined aftereffects in see-through video. Biocca and Roland (1998) showed that users would underreach for objects when wearing a see-through HMD, and that once the headset was taken off they would overcompensate by overreaching for the objects.[12] More recently, scholars have directly examined video passthrough and found that distances were underestimated in a blind throwing task in video passthrough compared to without a headset.[13] Those authors largely attributed this to the limited field-of-view of the headset. Similar results were found by Park and colleagues (2008) who showed significant decreases in spatial task accuracy in video see-through compared to without it.[14] However, to date there has been no quantitative research which examines the *aftereffects* of passthrough video. Therefore, we seek to understand how users experience differences in distance perception in video passthrough compared to the physical world, how these effects change with time in the HMD, and the aftereffects of HMD removal (**RQ3**).

Beyond the individual effects of passthrough, recent work has illustrated the influence of see-through VR use, ie. the Microsoft HoloLens, in social interactions.[15] Miller et al. (2019) found that users in a headset felt less social connectedness and less social presence to their interaction partners compared to those not using an AR headset in passthrough, a phenomenon scholars now refer to as *social absence*.[15, 1] Therefore, this study also tests how social absence differs in mixed reality compared to without headset in an experimenter-participant setting (**RQ4**).

**Materials and Methods**

This study was approved by the Institutional Review Board at the *[institution blinded]* (Protocol #: 66462).

*Hardware*

Participants used the Meta Quest 3 headset (515 g) with 2064 x 2208 pixels per eye, 110 x 96-degree field-of-view, 39 milliseconds latency, 120 Hz refresh rate, and full color passthrough. Hand controllers were not used during the study.

*Procedure*

The study employed a repeated measures design. Following their completion of the pre-experiment survey consisting of demographic questions and past VR and passthrough related experiences, participants were randomly assigned to a (1) *passthrough followed by no headset condition* or a (2) *no headset followed by passthrough condition* to counterbalance order effects. In both conditions, participants performed Basic Body Awareness Therapy, and the VR avatar body transfer exercises to elicit a feeling of body awareness and a sense of embodiment. These exercises invited participants to rotate their torso, wave at their mirror image, walk in place, and stand still (**Appendix A**). Each movement was conducted for 15-30 seconds in front of a mirror (**Figure 1**).[16-18] This was followed by a blind walk task to assess how accurately participants could estimate distances (**Figure 2**).[19] In total, participants wore the headset for 4 minutes for these exercises.

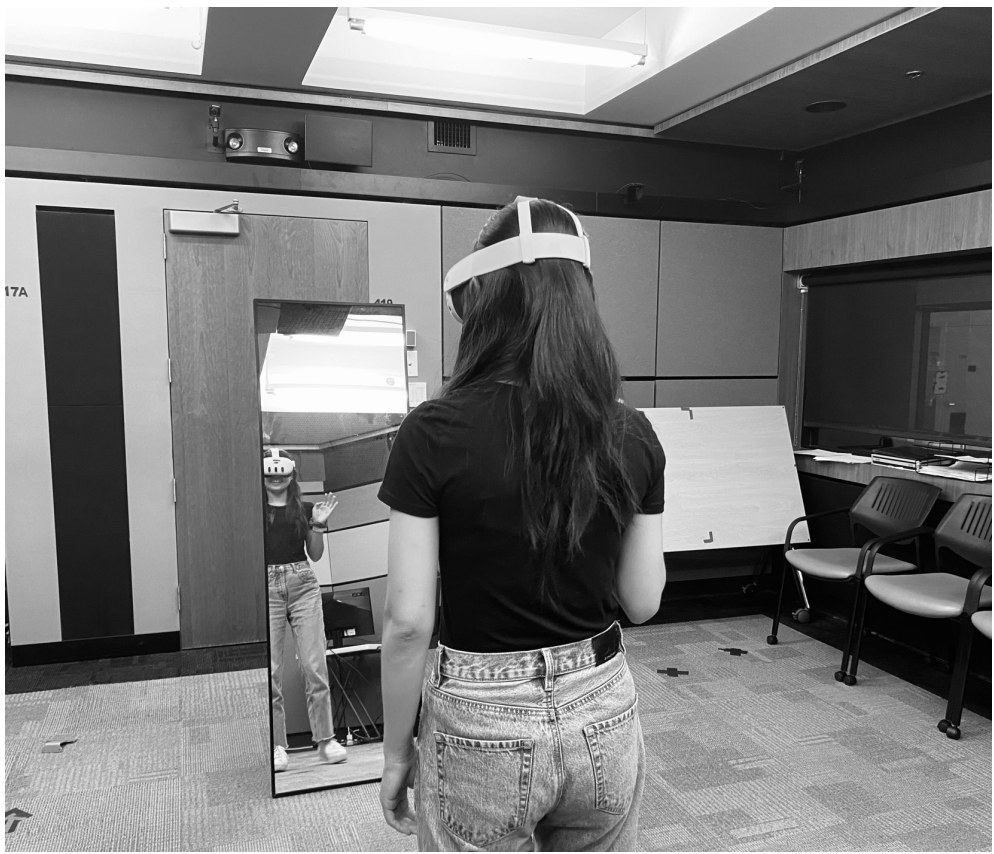

*Figure 1.* The participant wears an MR head-mounted display and participates in body transfer exercises for 4-minutes.

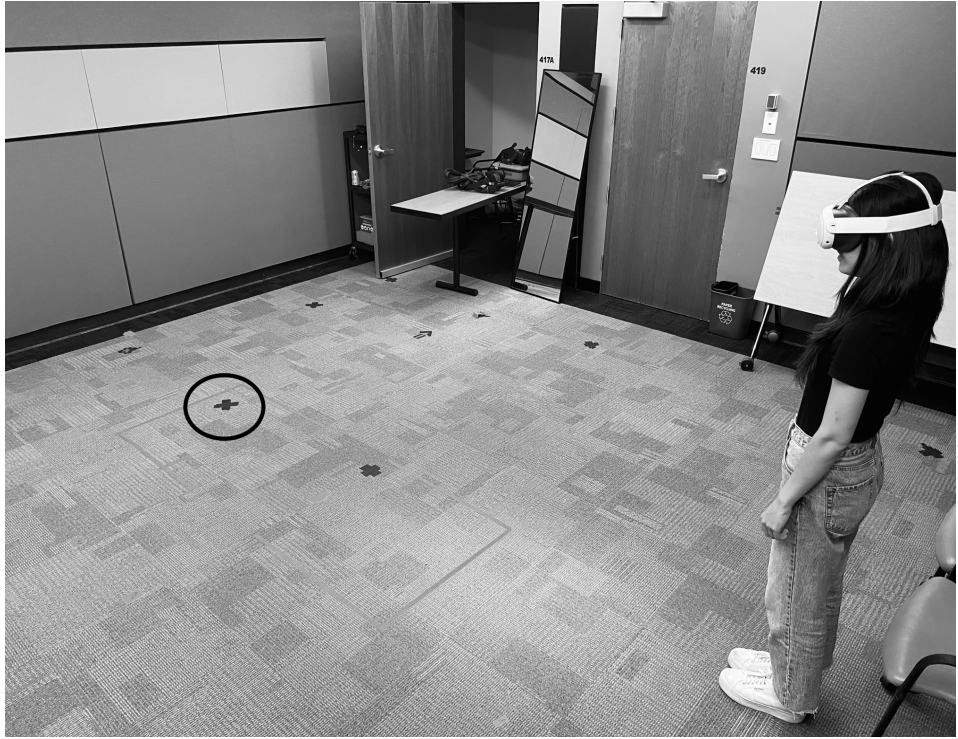

*Figure 2.* The participant wears an MR head-mounted display and engages in a blind walk task to the point circled that is 3.12 m away from them. They performed the blind walk task four times (i - in headset, ii - directly after removing headset, iii - without a headset, iv - without a headset for 3 more trials after initial no headset).

After completion of the first (passthrough or no passthrough) condition, participants completed a post-experiment questionnaire inquiring about simulator sickness, presence, embodiment, change in perceived body schema, and body distortions. They were then exposed to the second (passthrough or no passthrough) condition and performed the body transfer exercises; following which, they answered identical questions on simulator sickness, presence, embodiment, change in perceived body schema, and body distortions.

*Participants*

Prior to starting the experiment, we ran sample size calculation using G*Power with a power of 0.8, effect size of 0.5, and an alpha error probability of 0.05, which resulted in a power of 34 participants necessary for recruitment. We recruited participants via the *[institution blinded]* and online outreach. Participants were remunerated with a $25 gift card or granted course credit. A total of 40 participants consented to participate. Participants who did not consent to having their video recorded were filtered out of the video analysis portion of the study (N=3). Participant ages ranged from 19-37 (mean [M] = 23.15, standard deviation [SD] = 4.45). ***Table 1*** presents additional participant demographic information.

**Table 1.** *Demographic Characteristics of Participants (N=40)*

| Gender | N |
|---|---|
| Woman | 29 |
| Men | 10 |
| Decline to Answer | 1 |
| **Race/Ethnicity** | |
| African, African American, or Black | 5 |
| Asian or Asian-American | 13 |
| Hispanic or Latinx | 5 |
| Middle Eastern | 2 |
| White | 7 |
| Mixed | 6 |
| Decline to answer | 2 |
| **Prior Virtual Reality Experience** | |
| Never | 10 |
| Rarely | 27 |
| Sometimes | 2 |
| Several Times a Week | 1 |
| **Prior Passthrough Experience** | |
| Never | 26 |
| Rarely | 12 |
| Sometimes | 1 |
| Several Times a Week | 1 |

*Measures*

*Distance Perception.* Participants engaged in a blind walk task where they were instructed to walk toward a target they had seen previously in the lab with their eyes closed (i) 3 times in passthrough, (ii) 3 times directly following the use of passthrough, (iii) 3 times without headset, (iv) and 3 trials without headset after the initial three without headset trials. A researcher measured the difference in distance between where participants perceived the target to be and its actual location.[19]

*Simulator Sickness Questionnaire.* Simulator sickness was assessed using the Simulator Sickness Questionnaire (Cronbach's $\alpha$ = 0.87).[20,21] The questionnaire is made up of 3 subscales, namely, "Nausea," "Oculomotor," and "Disorientation," and 16-items on a 4-point scale (0 = "None" and 3 = "Severe"). Total scores can be associated with negligible (< 5), minimal (5-10), significant (10-15), and concerning (15-20), with above 20 being considered bad.[22]

*Embodiment Questionnaire.* Sense of embodiment was measured using an adapted avatar embodiment questionnaire (Cronbach's $\alpha$ = 0.89).[23] Of the 16-item 7-point scale (1 = "Never" and 7 = "Always"), we selected 9-items that were later modified to be suitable for the passthrough scenario (**Appendix B**).

*Change in Perceived Body Schema.* Change in perceived body schema, or the change that participants perceive and feel toward the spatial representation of their body, was measured using 6-items from the Virtual Embodiment Questionnaire (1 = "Strongly disagree" and 7 = "Strongly agree") (Cronbach's $\alpha$ = 0.76 – 0.78).[24]

*Body Distortion.* Body distortion was measured with four items on a 5-point scale where 1 = "Not at all" and 5 = "Extremely" (Cronbach's $\alpha$ = 0.79) (**Appendix C**).

*Presence.* Presence was measured using a 5-point (1 = "Not at all" and 5 = "Extremely") adapted Virtual Human Interaction Lab (VHIL) Presence scale (2022) with three subscales namely, social presence (Cronbach's $\alpha$ = 0.86), self-presence (Cronbach's $\alpha$ = 0.63) and spatial presence (Cronbach's $\alpha$ = 0.91) (**Appendix D**).[25]

**Results**

Means and standard deviations of all dependent variables by condition are shown in **Table 2**. We first used the Shapiro-Wilk normality test to assess data distribution across variables and conditions. As the data was non-parametric and not normally distributed, continuous outcome variables were analyzed using the Wilcoxon signed-rank test. To account for outliers in variables, we followed the winsorizing median statistical technique that computes the median of that variable and replaces outliers with the observations closest to them. Findings were the same with and without the statistical transformation of outliers. For each test reported below we did not report condition order as an independent variable as it did not change the pattern of results.

*Table 2.* *Means and Standard Deviation Table of Dependent Variables in the Passthrough and No Headset Condition (N=40)*[a]

| Variables | In Passthrough – Means (SD) | No Headset – Means (SD) |
|---|---|---|
| **Embodiment** | 3.15 (1.35) | 2.96 (1.55) |
| Appearance | 2.67 (1.33) | 2.71 (1.46) |
| Response | 2.70 (1.60) | 2.24 (1.55)* |
| Ownership | 3.87 (1.42) | 3.67 (1.84) |
| Multi-Sensory | 3.33 (1.64) | 3.29 (1.93) |
| **Perceived Change in Body Schema** | 3.06 (1.53) | 2.21 (1.28)* |
| **Body Distortion** | 2.01 (0.95) | 1.13 (0.28)*** |
| **Presence** | | |
| Social Presence | 2.48 (0.99) | 3.08 (1.22)*** |
| Self-Presence | 3.62 (0.83) | 3.90 (0.91) |
| Spatial Presence | 2.51 (1.18) | 1.36 (0.70)*** |
| **Simulator Sickness** | | |
| Nausea | 12.48 (12.41) | 3.93 (4.84)* |
| Oculomotor | 20.60 (17.39) | 6.89 (7.73)** |
| Disorientation | 45.06 (38.62) | 27.84 (27.84)*** |
| Total Simulator Sickness | 26.87 (22.22) | 9.44 (9.21)** |

[a] ***$p < 0.001$; **$p < 0.01$; *$p < 0.05$

### *Passthrough, even in short doses, can create simulator sickness*

With reference to our first research question, when comparing simulator sickness between participants following video passthrough and no headset, there was a significant difference. A Wilcoxon signed-rank test was used to compare each simulator sickness category. Significant findings were found when comparing each category. Nausea in passthrough (M = 12.48, SD = 12.41) was significantly higher than without a headset (M = 3.93, SD = 4.84), (W = 458.50, $p < 0.05$). Oculomotor in passthrough (M = 20.60, SD = 17.39) was significantly higher than without a headset (M = 6.89, SD = 7.73), (W = 644, $p < 0.05$). Disorientation in passthrough (M = 45.06, SD = 38.62) was significantly higher than without a headset (M = 27.84, SD = 27.84), (W = 606.5, $p < 0.0001$). Total simulator sickness in passthrough (M = 26.87, SD = 22.22) was significantly higher than without a headset (M = 9.44, SD = 9.21), (W = 587, $p <$

0.01). These findings indicate that passthrough, even in short doses, can create simulator sickness (**Figure 3**).

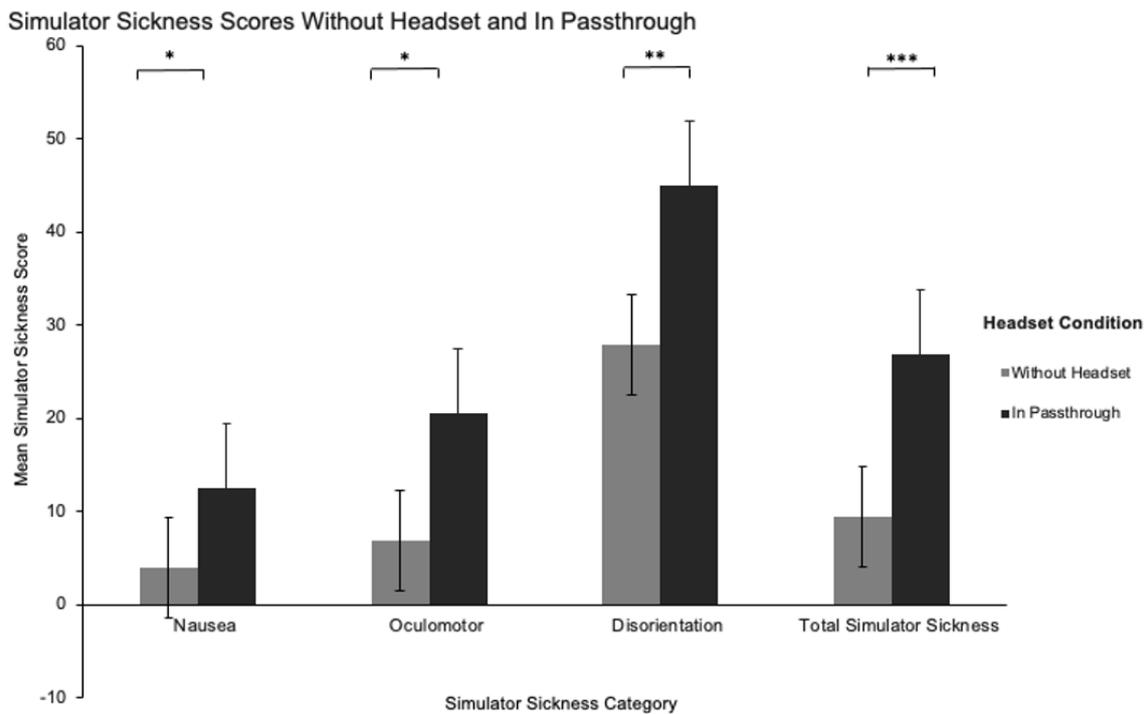

*Figure 3.* A bar plot illustrating simulator sickness scores in passthrough and without a headset by simulator sickness categories.

***Participants experienced perceived change in body schema and body distortions in video passthrough***

With reference to our second research question, we ran a Wilcoxon signed-rank test which indicated no significant differences between participants' sense of embodiment in video passthrough or the physical environment ($z = 1.27$, *n.*s), however, participants did experience a significantly higher change in their perceived body schema in video passthrough ($M = 3.06$, $SD = 1.53$) compared to without headset ($M = 2.21$, $SD = 1.28$). Participants also reported experiencing significantly higher body distortions in video passthrough ($M = 2.01$, $SD = 0.95$) compared to without headset ($M = 1.13$, $SD = 0.28$) (**Table 2**).

***Underestimation of distance occurs in video passthrough compared to without headset and directly following video passthrough experience***

Considering our third research question, a linear mixed effects model was used to account for trial time as a continuous variable, since each participant did 3 trials per headset trial type (in passthrough, directly following passthrough, no headset, and no headset following initial 3 trials) and condition order in which HMD was used. There were no changes in the pattern of results by trial time and order.

Participants exhibited significant underestimation of distances while in passthrough compared to the three blocks where they estimated distance without a headset (the first three trials and the second three trials from the without condition, and the second block after wearing the headset in the headset condition). Comparing distance underestimation in passthrough and directly following passthrough (participants remove their headset directly after using it), the average participant who just removed their headset underestimated distance less by 21.27 cm than in passthrough (SE = 3.48 cm) ($p < 0.001$) (**Figure 4**).Comparing the amount of distance underestimation in passthrough and no headset (the two blocks where participants did not use headset at all), we found that there were significant differences. In comparison to wearing passthrough, the average participant who did not wear a headset underestimated distances less by 8.85 cm (SE = 3.54 cm) ($p < 0.05$) (**Figure 4**).

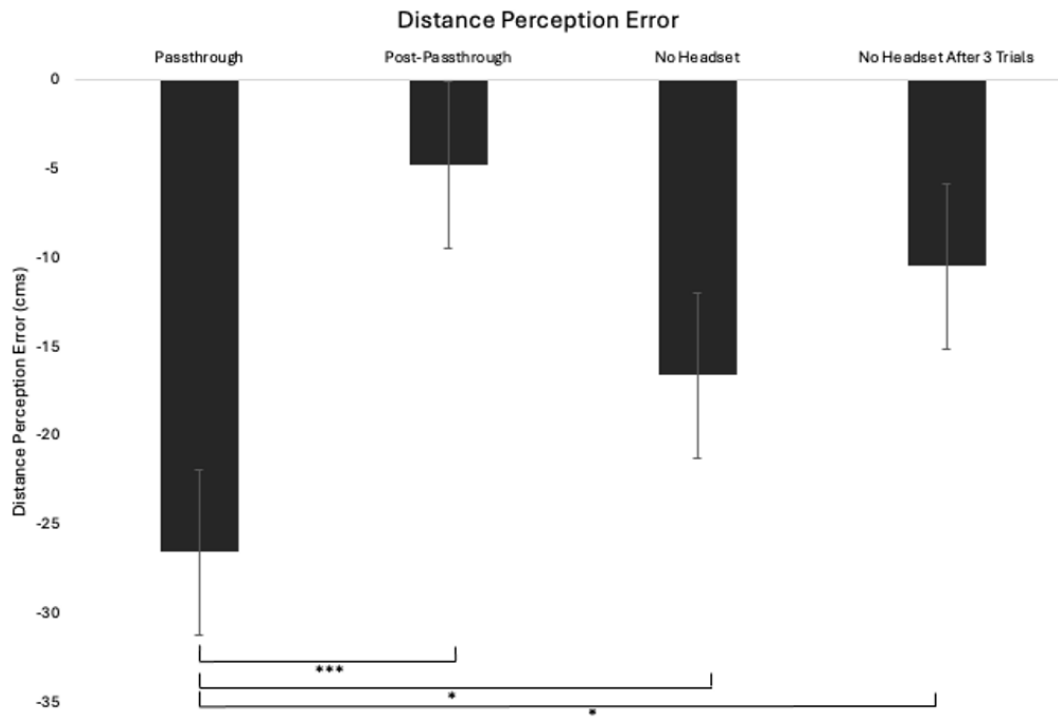

*Figure 4.* Differences in distance underestimation in video passthrough and without headset.

*Passthrough causes individuals to feel socially absent*

To answer our fourth research question that explores the relationship between social presence scores and without a headset and in passthrough, we ran a Wilcoxon signed rank test. The results showed that without a headset (M = 2.48, SD = 0.99), social presence was higher than in video passthrough (M = 3.08, SD = 1.22), (W = 17.50, p < 0.001; **Figure 5**). Median values show an increase of 30.55% in social presence ratings without wearing the headset compared to in video passthrough.

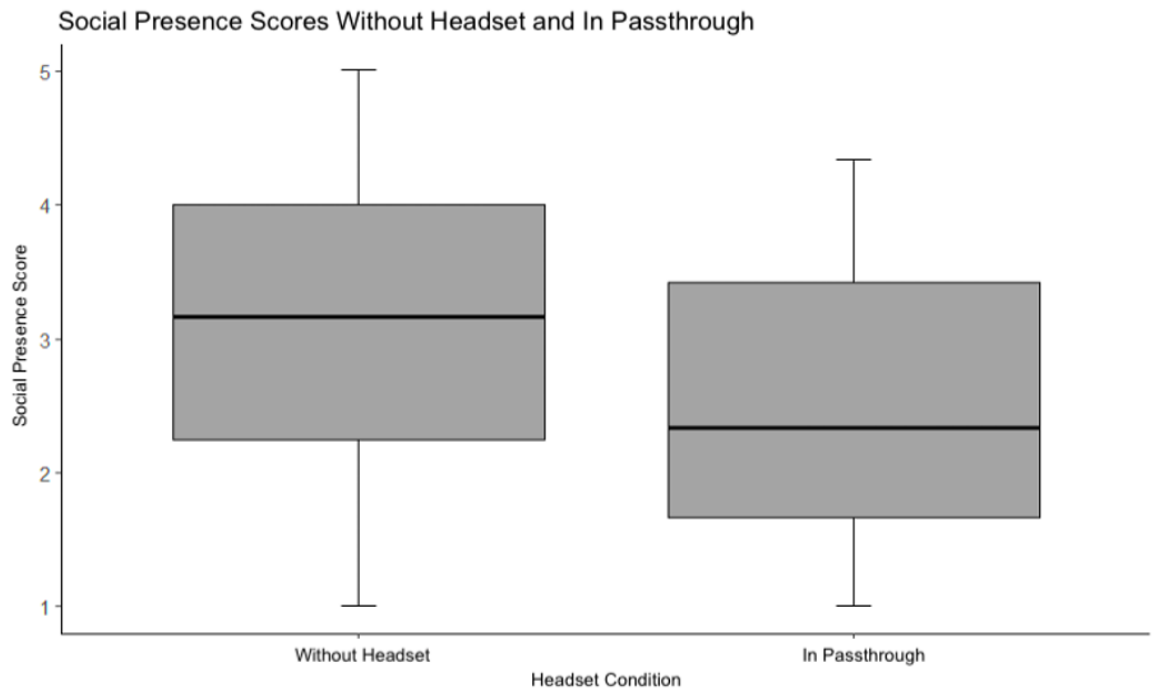

*Figure 5.* A box plot illustrating the difference between mean social presence scores in two conditions - without headset and in passthrough.

**Discussion**

While prior literature has shed light on the influence of MR content on user psychology,[8,26-27] few have directly compared face-to-face interaction to the video passthrough as a medium. This study draws from psychology and communication theories, as well as device affordances, to explore and understand the psychological and behavioral effects of video passthrough.

The body transfer exercises that participants performed early on in our study procedure was aimed to induce a sense of embodiment and body awareness.[17-18] Following these exercises, our findings indicated that while there were no differences in participants' embodiment in their physical bodies both in passthrough and without headset. However, there was a change in users' perceived body schema and body distortion while in passthrough compared to the physical reality. This could be a result of the curvature of the small screens in the headset and the algorithmic process of integrating multiple camera streams which cause a discrepancy in the location of the user's real eyes and the camera display's location.[5] These distortions create changes in body perceptions and should be given attention, particularly given the growing number of mixed reality interventions on body image.[28-29] This finding generalizes past looking in mirrors as individuals are looking down at their arms and torsos often while wearing video passthrough.

Moreover, our findings shed light on how even in short doses of 4 minutes and without any virtual effects layered over the real world, users feel simulator sickness, with disorientation being the highest contributor to sickness. We also extend prior literature in support of sensory conflict theory, suggesting that MR headsets give rise to incongruencies of conflicting information and desynchronized sensory outputs, which may be attributed to technical features such as lower refresh rates and resolution compared to the human eye.[30] However, while our work builds on the precedent that simulator sickness is zero before headset use,[20,31-32] another line of work has also illustrated that the non-zero baseline may be incorrect.[33-34] As such, future research should aim to compare simulator sickness in video passthrough and non-zero headset baselines, as well as whether seeing themselves in a mirror influenced levels of sickness.

With reference to the perceptual effects of distance underestimation in passthrough, our findings implore the need for caution in the use of video passthrough in motion due to higher distance underestimation while in headset compared to without it. This underestimation of distance in MR headsets is a well-documented finding that we replicate and can be attributed to the weight and field-of-view of video passthrough as well as the geometric distortions in the display graphics.[35, 13, 36, 37,38] We also find that while participants underestimated distances in passthrough, there were no significant aftereffects from a very short headset experience.

Finally, on exploring the social implications of MR headsets as they relate to social absence, we further illustrated that these diminished social presence in the physical world show how users may treat being in passthrough the same way they treat being in VR, as though they were psychologically removed from the real world and immersed in a virtual world.[39-40] In essence, passthrough video turns a face to face interaction into a Zoom call, where real people become mediated by video. Possible explanations for this could include the low field-of-view in a headset compared to without one and the lower visual fidelity of the physical surrounding in passthrough. Further investigation on social presence in MR can also shed light on the possibility of social facilitation and inhibition processes when in passthrough.[15, 41,42] Given that our findings illustrate *social absence* only as it relates to the experimenter and passthrough user, future work should seek to explore how users experience this phenomenon in dyadic and group interactions and test out the mechanisms under which social facilitation and inhibition processes may occur.[15]

**Conclusion**

This study provides initial quantitative evidence that even with the advances of video passthrough in the most recent MR headsets, perceptual and social effects will limit the headsets from becoming an everyday medium that augments physical world interactions. Our work underscores the need for further research into the longitudinal and long-term effects of video passthrough immersion to better understand how physiological indicators such as simulator sickness and psychological factors such as social absence can be minimized.

**Supplemental Files**

**Appendix A. Instructions for Body Transfer Task.**
We used the body transfer task to elicit a feeling of body awareness and a sense of embodiment. These were adapted from Waltemate et al. (2018), Gyllensten et al. (2018), and Dollinger et al. (2022). The instructions given to participants are listed below.

Set 1 (Each exercise was performed for 20 seconds each):
1. Lift your right arm and wave to your mirror image in a relaxed way.
2. Now wave your other hand.
3. Walk in place and lift your knees as high as your hips.
4. Stretch out both arms to the front and perform circular movements.
5. Stretch out your right arm to the side and perform circular movements.

Set 2 (Each exercise was performed for 30 seconds each):
1. Stretch out your left arm to the side and perform circular movements.
2. Stand still and focus on perceiving your posture.
3. Rotate your torso.
4. Perform rocking movements with your legs while letting your arm swing.
5. Push your hands forwards while standing in a step position.
6. Rock your legs and swing your arms.

**Appendix B. Adapted Embodiment Questionnaire.**

Sense of embodiment was measured using an adapted avatar embodiment questionnaire (Gonzalez-Franco & Peck, 2021). Of the 16-item 7-point scale, where 1 = "Never", 2 = "Almost Never", 3 = "Rarely", 4 = "Half of the time", 5 = "Often", 6 = "Most of the time" and 7 = "Always", we selected 9-items that were later modified to be suitable for the passthrough scenario as shown below:
1. I felt out of my body.
2. I felt as if my (real) body were drifting toward the mirror digital self-representation.
3. I felt as if the movements of the mirror digital self-representation were influencing my own movements.
4. It felt as if my (real) body were turning into my mirror digital self-representation.
5. At some point it felt as if my real body was starting to take on the posture or shape of the mirror digital self-representation that I saw.
6. I felt as if my body had changed.
7. At some point it felt that the mirror digital self-representation resembled my own (real) body, in terms of shape, skin tone or other visual features.
8. I felt as if my body was located where I saw the mirror digital self-representation.
9. I felt as if the mirror digital self-representation were drifting toward my (real) body.

**Appendix C. Body Distortion Scale.**

Body distortion was measured using a 4-item 5-point scale questionnaire where 1 = "Not at all", 2 = "Slightly", 3 = "Moderately", 4 = "Very" and 5 = "Extremely". To ensure that the scale we developed for this study accurately measured this construct, we conducted an exploratory factor analysis indicating $\alpha = 0.79$. The items included:
1. I experienced the sensation of my limbs being attached to my body in the wrong locations.

2. I experienced my body curving in weird ways.
3. The displayed passthrough image of my body in the mirror did not match what I would see without the headset.
4. The distortion of my body in passthrough changed the way I felt about my body.

**Appendix D. Presence.**

Presence was measured using the adapted Virtual Human Interaction Lab (VHIL) Presence scale (2022) where social, self, and spatial presence was measured using a 5-point Likert scale where 1 = "Not at all", 2 = "Slightly", 3 = "Moderately", 4 = "Very", 5 = "Extremely."
1. It felt like the experimenter was in the room with me.
2. It felt like I was face-to-face with the experimenter.
3. It felt like the experimenter was aware of my presence.
4. I felt that my digital mirror self-representation represented me.
5. When something happened to my digital mirror self-representation, I felt like it was happening to me.
6. I felt like I was able to control my digital mirror self-representation as though it were my own.
7. It felt as if I was inside the virtual world.
8. It felt as if I was visiting another place.